\documentclass[aps,pre,amsmath,superscriptaddress,amssymb,reprint,floatfix]{revtex4-1}     
\usepackage{graphics}
\usepackage{color}
\usepackage{hyperref} 
\usepackage{epstopdf}
\usepackage{ifpdf} 
\usepackage{epsfig}

\begin{document}
\title{Mapping diffusivity of narrow channels into one-dimension}

\author{Mahdi Zarif}
\affiliation{Department of Physical and Computational Chemistry, Shahid Beheshti University, Tehran 19839-9411, Iran}
\email{m\_zarif@sbu.ac.ir}

\author{Richard K. Bowles}
\affiliation{Department of Chemistry, University of Saskatchewan, Saskatoon, S7N 5C9, Canada}
\email{richard.bowles@usask.ca}

\begin{abstract}
The diffusion of particles trapped in long narrow channels occurs predominantly in one dimension. Here, molecular dynamics simulation is used to study the inertial dynamics of two-dimensional hard disks, confined to long, narrow, structureless channels with hard walls in the no-passing regime. We show that the diffusion coefficient obtained from the mean squared displacement can be mapped onto the exact results for the diffusion of the strictly one-dimensional hard rod system through an effective occupied volume fraction obtained from either the equation of state or a geometric projection of the particle interaction diameters.
\end{abstract}
\date{revised version -- \today}
\maketitle

\section{Introduction}
\label{sec:intro}
With recent advances in nanotechnology and our ability to produce nano-structured materials, there is now considerable interest in understanding the transport of fluids inside them~\cite{Robertson:PNAS2006,Gubbins2011,Rodr:EPJE2012}. Fluids in highly confined systems such as carbon nanotubes~\cite{Mukh:2010}, zeolites~\cite{karger1992,kumar2014,karger2015}, metal--organic frameworks~\cite{Jobic2016} and ion channels in biological membranes~\cite{Jensen2002,Rasaiah2008} exhibit single-file diffusion (SFD) in which particles cannot pass each other and the diffusion is constrained in one dimension. This geometric restriction, combined with the nature of the fundamental particles dynamics, can have a significant effect on the way the particles move. For example, the transport coefficient of a fluid along an arbitrary axis, $D_x$, can generally be calculated from the long time behavior of the mean squared displacement (MSD), which is given by the Einstein relation:
\begin{equation}
\left \langle \left (x(t)-x(0)\right )^{2} \right \rangle = \left \langle \Delta x^{2}\left (t\right ) \right \rangle \propto \alpha l (D_xt)^{\gamma} ,
\label{eq:eqn1}
\end{equation}
where $\alpha$ depends upon the distribution of jumps in the basic motion,  $l$ is the ``free volume" along the $x$ axis per particle and $\gamma$ is a variable~\cite{Levitt:1973p14144}. In a fluid performing deterministic dynamics, the particles exhibit ballistic motion at short times, where  $\gamma=2$, before caging and cage hopping leads to a crossover to diffusive motion at long times where  $\gamma=1$, and Eq.~\ref{eq:eqn1} becomes,
\begin{equation}
\left \langle \Delta x^{2}\left (t\right ) \right \rangle=2D_{x}t \mbox{.}\\
\label{eq:msdbulk}
\end{equation}
Equation~\ref{eq:msdbulk} is valid in the bulk at long times, independent of the nature of the particle dynamics, and remains true in single file systems when the dynamics is deterministic~\cite{Hahn_Molecular_1996}. However, if the single file particles are subject to a Brownian background~\cite{Levitt:1973p14144}, or independent stochastic forces~\cite{precus:1974}, then the MSD of a tagged particle is described by an Einstein-like relation,
\begin{equation}
\left \langle \Delta x^{2}\left (t\right ) \right \rangle=2F_{x}t^{1/2}\mbox{ ,}\\
\label{eq:F}
\end{equation}
where $\gamma=1/2$, indicating the system exhibits anomalous sub-diffusion, and the diffusion coefficient has been replaced by the mobility factor $F_x$~\cite{hahn:1998,lin:2005}.

One of the most challenging tasks in the field of condensed matter is to find a relationship between the transport coefficients of a fluid and its equilibrium thermodynamics properties. Rosenfeld~\cite{Rosenfeld:1977p14198} originally proposed a scaling relation, $D_{\textup{R}} \approx 0.58 \; \textup{exp}(As^{\textup{ex}})$ between a reduced transport coefficient, $D_{\textup{R}}=DT^{-1/2}\rho^{1/3}$ ($\rho$ is number density), and the excess entropy per particle, $s^{\textup{ex}}$, where $A$ is a material dependent constant. This was later expanded to consider dilute fluids~\cite{Rosenfeld:1999p14199}. Dzugutov~\cite{Dzugutov:1996p14200} developed a similar scaling relation, $D_{\textup{D}} \approx 0.078 \; \textup{exp}(s^{\textup{ex}})$, linking the excess entropy to a diffusion coefficient, $D_{\textup{D}}=D\rho^{2/3}\Gamma_{\textup{E}}^{-1}$, that is reduced in terms of an effective Enskog inter-particle collision frequency, $\Gamma_E$, that includes contributions obtained from microscopic structural quantities such as the radial distribution function. These studies have subsequently formed the basis for scaling type approaches focused on connecting the excess entropy to dynamics in a variety of systems~\cite{Bretonnet:2002p14205,yokoyama:HTMP2011,cao:PBCM2011,cao:JAP2015,cao:RSC2016,Jakse:SR2016,samanta:PRL2004,wautelet:EJP2001,colbrook:MNRAS2017,Gioia:BLM2004,BOMONT:PA2015,Li:CPL2004,Gompper:EPJE2011,binder:EPJE2009}. Furthermore, it is important to note that entropic scaling laws, such as the Adam-Gibbs relation~\cite{adam:1965}, have also successfully described diffusion and structural relaxation in terms of the configurational entropy in supercooled liquids~\cite{sastry:2001,speedy:2001}, at least at temperatures above the observed break down in the Stokes-Einstein relation~\cite{sastry:EPJE2018}.

In the context of confined fluids, Mittal et. al.~\cite{Mittal:2006p14186} showed that the scaling relationships between the self-diffusivity, excess entropy and density remain valid for both quasi-two dimensional (particles confined between plates) and quasi-one-dimensional fluids. The same authors also found that self diffusivity of the confined systems could essentially be mapped onto the behaviour of bulk fluids through either an effective packing fraction or the excess entropy~\cite{Mittal:2007p14189}. Similarly, it has been shown that a slightly modified version of the Dzugutov scaling could account for the diffusivity in water and water-methane mixtures both in the bulk and when confined to carbon nanotubes of different diameters~\cite{sahu:jpcc2017}. This suggests that the properties of bulk materials could be used to predict the dynamics of highly confined fluids.

However, rather than use the bulk as a reference, Percus and co-workers~\cite{kalinay:2005jcp,percus:2010jsp,kalinay2011pre} have explored the possibility of mapping the dynamic properties of quasi-one-dimensional fluids to those of a truly one-dimensional ($1d$) system where it is often possible to obtain exact results, or is relatively easy to perform highly accurate numerical simulations. In particular, Percus~\cite{percus:2010jsp} developed a heuristic approach that suggests,
\begin{equation}
\left \langle \Delta x^{2}\left (t\right ) \right \rangle = K \left<\left|\Delta x(t)\right|\right>_0\mbox{,}\\
\label{eq:percus}
\end{equation}
where $\left<\left|\Delta x(t)\right|\right>_0$ contains the details of the particle motion in the strictly one-dimensional system, so is proportional to $t$ for inertial dynamics and proportional to $t^{1/2}$ in the case of anomalous sub-diffusion. For a $1d$ system of $N$ hard rods with diameter, $\sigma$, on a line of length $L$,  $K=(L/N-\sigma)$, but more generally $K$ could be approximated by $kT/ P$, or $(-kT\partial L/\partial NP)^{1/2}$, where the $P$ is the pressure in the quasi-$1d$ system, $T$ is the temperature and $k$ is the Boltzmann constant. The proposed strategy is then to identify characteristics of the distribution of next nearest neighbour separations, to which $P$ is connected, that can be used to map the dynamics between the system of interest and the $1d$ reference system.

The goal of the current paper is to simply demonstrate that an effective density and effective particle diameter, obtained either from the equation of state or a direct geometric measure, is able to provide such a map between the long time dynamics of a strictly $1d$ hard rod system and the dynamics of a quasi-$1d$ system. Here, we will focus on the study of two dimensional hard discs confined to a structureless channel exhibiting deterministic dynamics as an example. The remainder of the paper is organized as follows: Section~\ref{sec:method} describes the details of the model and the simulation method used to examine the dynamics. Section~\ref{sec:scaling} outlines the calculation of the effective density and particle size, and the mapping of the dynamics of the $1d$ hard rod system. We also show that the same scaling appears through the use of the excess entropy. Our conclusions are summarized in Section~\ref{sec:con}.

\section{Model and Methods}
\label{sec:method}

We consider a system of two-dimensional hard disks of diameter, $\sigma$, confined between parallel hard lines separated by a distance $H_d$. The channel length in the longitudinal direction is $L$ and the two ends obey periodic boundary conditions. The occupied volume fraction for the system is then given by,
\begin{equation}
\phi = \frac{N \pi \sigma ^{2}}{4H_{d}L}\mbox{,}\\
\label{eq:2dphi}
\end{equation}
where $N$ is the number of particles. The particle--particle and particle--wall interaction potentials are given by,
\begin{equation}
V(r_{ij})= \left\{\begin{matrix}
0 \;\;\; & r_{ij} \geqslant \sigma \\
\infty \;\;\; & r_{ij} <  \sigma
\end{matrix}\right.,
\label{eq:potential}
\end{equation}
\begin {equation}
V_{w}(r_{i})= \left\{\begin{matrix}
0 \;\;\; & \left | r_{y} \right | \leqslant h_{0}/2 \\
\infty \;\;\; & \textup{otherwise}
\end{matrix}\right.,
\label{eq:wpotential}
\end{equation}           
respectively, where $r_{ij}=\left | \mathbf{r_{j}-r_{i}} \right |$ is the distance between particles, $r_{y}$ is the component of the position vector for a particle perpendicular to the wall and $h_{0}=(H_{d}-\sigma)$, is the reduced channel diameter.

In the current work, we restrict our analysis to systems performing inertial dynamics, so in the long time limit the diffusion coefficient is given by Eq.~\ref{eq:msdbulk}. The MSD is obtained from event-driven molecular dynamics (EDMD) simulations~\cite{Alder:1960p258} in the canonical ($N,V,T$) ensemble  where $N=30000$ particles, $V=h_{0}L$ is the total available volume of system accessible to the center of the particles and $T$ is the temperature. The units of time in the simulation are $\sigma \sqrt {m/kT}$, where $m$ is the mass of the particles, which is set  equal to unity, and all lengths are in units of $\sigma$. We study channels with widths $H_{d} = 1.1$ up to ($1+\sqrt {3/4}$) which ensures that only nearest neighbors can interact and prevents the particles from passing each other.

At the start of each run, the particles are placed on a lattice at packing fraction $\phi = 0.1$ and are assigned a random distribution of velocities scaled to give $kT = 1$. At each $\phi$ studied, $200N$ collisions are used to reach equilibrium and the MSD in the longitudinal direction of the channel ($x$-axis) was measured over the next $400N$ collisions in which particle coordinates were saved 80 times, separated by $5N$ collisions. After collecting data at a given $\phi$ the system is compressed to a higher occupied volume fraction using a modified version of the Lubachevsky and Stillinger~\cite{Lubachevsky:1990p309} (LS) algorithm that ensures $H_{d}/\sigma$ remains constant as the diameter of the discs is changed ($L$ fixed), with a compression rate $ds= d\sigma/dt = 0.001$. The mean squared displacements are reported as averages over 20 independent runs performed for each channel diameter.

\begin{figure}
\includegraphics[width=1.0\linewidth]{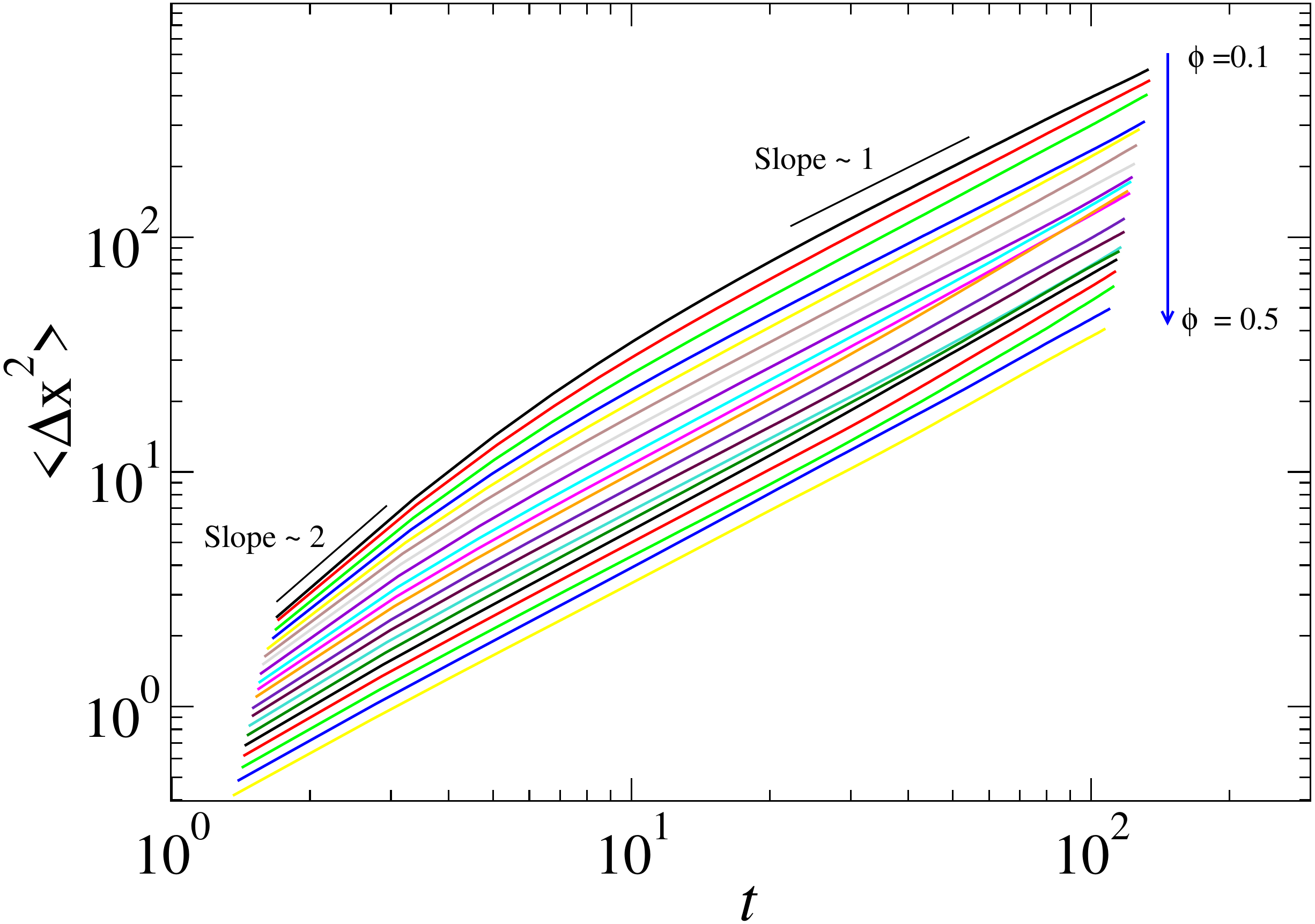}
\caption{The MSD along the direction of the pore axis, as a function of $t$ for $H_{d}=1.1$ at different $\phi$. The short line segments have slopes as indicated and are included as a guide to the eye.}
\label{fig:figure1}
\end{figure}

\begin{figure}
\includegraphics[width=1.0\linewidth]{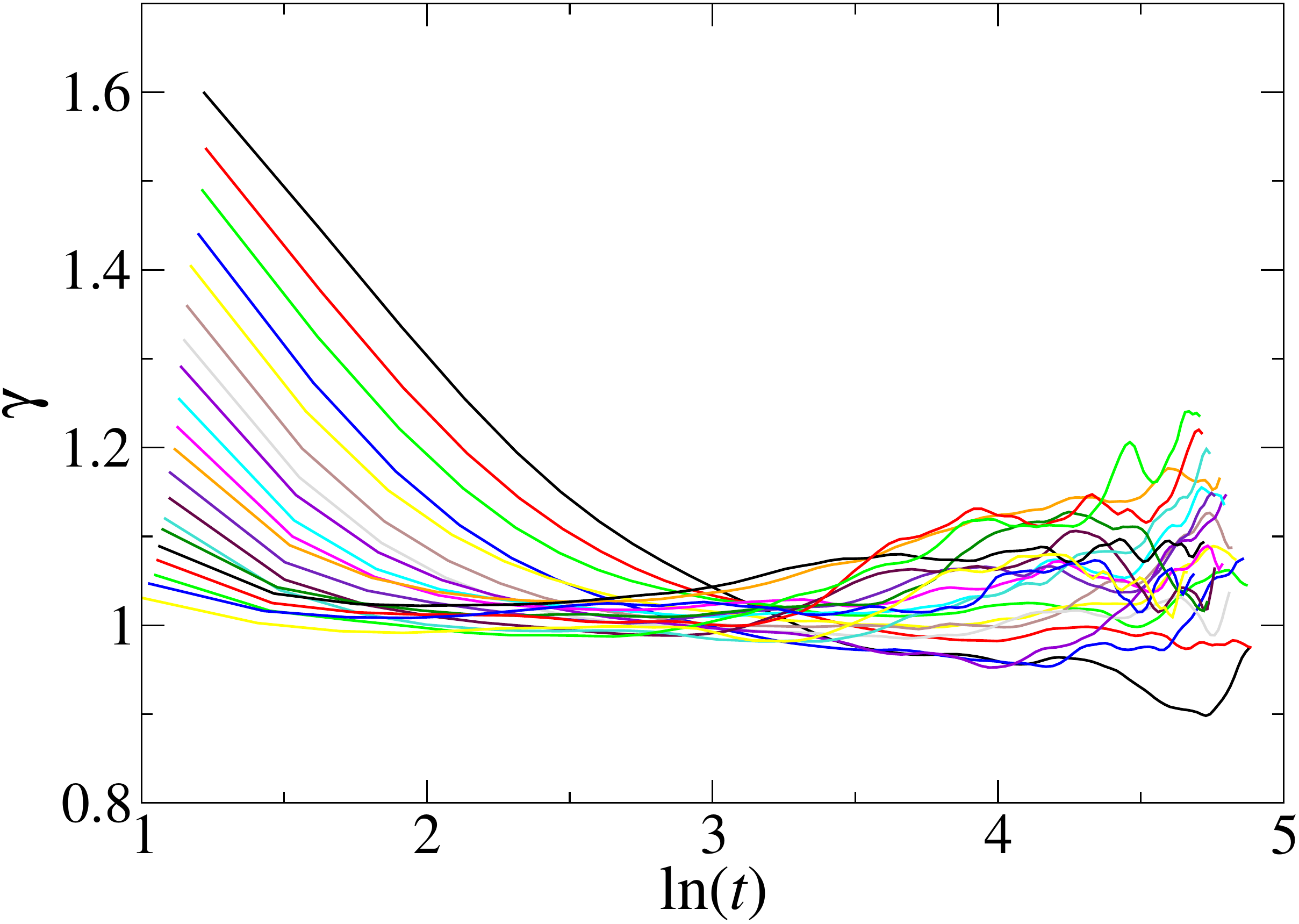}
\caption{Exponent, $\gamma$, as a function of $\ln t$ for $H_{d}=1.1$ at different $\phi$.}
\label{fig:figure2}
\end{figure}

\begin{figure}[t!]
\centering
\includegraphics[width=1.0\linewidth]{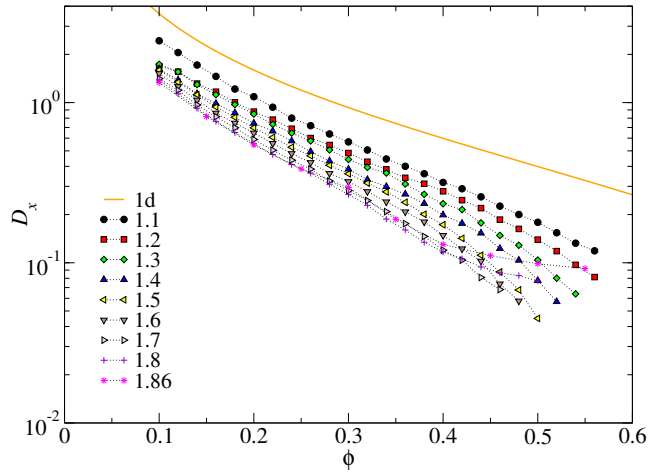}
\caption{Diffusivity, $D_x$ as a function of occupied volume fraction, $\phi$, for various channel widths.}
\label{fig:figure3}
\end{figure}

We monitor the self-diffusivity of the fluid by fitting the long-time behavior of the MSD in the longitudinal direction of the channel ($x$-axis) for the particles to the Einstein equation (Eq.~\ref{eq:eqn1}). Figure 1 shows the MSD as a function of time for the case with $H_{d} = 1.1$ at different densities (starting from $\phi =0.1$ up to $\phi = 0.5$ with steps of 0.02). We then obtain $\gamma$ by taking the numerical derivative of the data presented in Fig.~\ref{fig:figure1} to determine where we should evaluate the slope to obtain the diffusion coefficient. Figure~\ref{fig:figure2} shows $\gamma$ evolves over time. At low densities, $\gamma$ decreases slowly from its ballistic value because the particles are well separated. At higher densities, the expected diffusive behaviour is obtained quickly, but at longer times, correlations build up as a result of finite size effects~\cite{Hahn_Molecular_1996} and a non-monotonous behavior is observed. Figure~\ref{fig:figure3} shows the value of $D_x$, obtained by measuring the slope of the MSD over the time period where $\gamma\approx1$, as a function of occupied volume fraction for different channel diameters.

\begin{figure}
\includegraphics[width=1.0\linewidth]{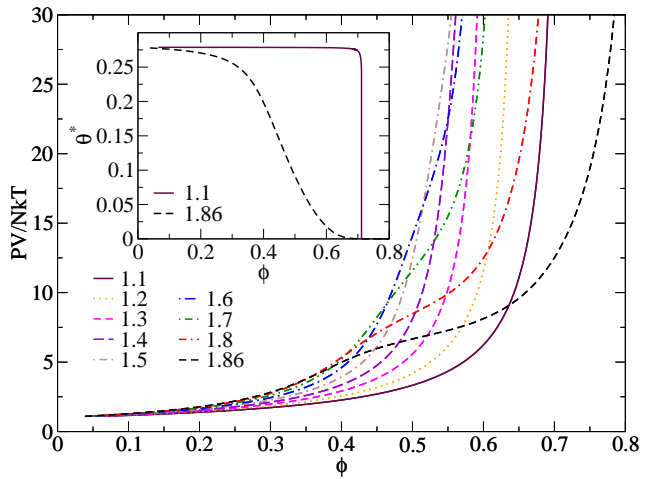}
\caption{$PV/NkT$ vs. $\phi$ for different channel $H_d$, obtained using the transfer matrix method described in Ref~\cite{Kofke:1993p3}. Inset: The equilibrium  fraction of defects, $\theta^*$, in the jammed states associated with the inherent structure basins sampled by the equilibrium fluid as a function $\phi$ for channel widths $H_d=1.1$ (solid line) and $H_d=1.86$ (dashed line).}
\label{fig:EOS}
\end{figure}

The equation of state for this quasi-$1d$ system is obtained following the transfer matrix analysis developed by Kofke and Post~\cite{Kofke:1993p3} and is shown in Figure~\ref{fig:EOS}.

\section{Scaling Dynamics with a One-Dimensional Effective Density and Particle Size}
\label{sec:scaling}

The exact equation for diffusion of hard rods in a strictly one-dimensional ($1d$) system was solved by Jepsen~\cite{Jepsen:1965p14139} and is given by:
\begin{equation}
\frac{D_x}{\sigma} = \frac{(1-\phi)}{\phi (2 \pi \beta m)^{1/2}},
\label{eq:eqn2}
\end{equation}
where $\beta=(kT)^{-1}$, $m$ is the mass of a particle, and the occupied volume fraction becomes $\phi = N\sigma/L$ in $1d$. Figure~\ref{fig:figure3} compares $D_x$ for the quasi-$1d$ systems with the strictly $1d$ system as a function of occupied volume fraction for different channel diameters. At low densities, $D_x$ for the quasi-one-dimensional channels generally follows the trend of the strictly $1d$ system, but decreases with increasing $H_d$ at fixed $\phi$. At higher densities, the diffusion coefficients begin to exhibit behaviour that reflects the differences between the two. For the narrower quasi-$1d$ channels, $D_x$ decreases more rapidly than the strictly $1d$ system because $D_x$ necessary goes to zero as $\phi$ approaches $\phi_{J_{max}}<1$, the most dense packing for the system. For the wider channels, $D_x$ exhibits a plateau at intermediate densities that coincides with the plateau observed in the equation of state (see Fig~\ref{fig:EOS}). While the short ranged hard disc interaction and quasi-one-dimensional nature of the system rule out the possibility that the system has a phase transition, the fluid undergoes a significant, but continuous, structural rearrangement at intermediate densities~\cite{Kofke:1993p3,Bowles_Landscapes_2006}. The low density fluid is characterized by linear arrangements of the discs where the collisions between discs occur predominantly in the longitudinal direction, and the high density fluid is dominated by zig-zag arrangements of the discs, with disc interactions occurring across the channel. In the transition region, the effects of increasing $\phi$ are somewhat off set by these structural rearrangements so the amount of space the discs have to move in the longitudinal direction changes slowly, giving rise to the plateaus in the diffusion coefficient and the longitudinal pressure. The diffusion coefficient would be expected to decrease again at higher densities where the pressure begins to increase again, but our simulations becomes trapped in a glassy state where it is not possible to measure the equilibrium properties of the fluid. 

The goal of this work is to identify a scaling that accounts for these differences and collapses $D_x$ for the quasi-$1d$ systems onto Eq.~\ref{eq:eqn2}. As the diameter of channel changes, the kinetic term in Eq.~\ref{eq:eqn2}, $(2\pi\beta m)^{-1/2}$ remains unchanged, so the focus of our task is to provide a map between the quasi-$1d$ and strictly-$1d$ systems for the density dependent term, $(1-\phi)/\phi$, which can be thought of as the mean clearance between neighbouring particles~\cite{Hahn_Molecular_1996}.  One possibility is to explore this term in the context of the inherent structure paradigm~\cite{Stillinger_Systematic_1964,Stillinger_Hidden_1982}, where configurations that map to the same mechanically stable jammed structure, or inherent structure, are grouped together into local basins of attraction so that the thermodynamics and dynamics of the fluid can be described in terms of the basins sampled at equilibrium. The amount of vibrational space the particles have when sampling a given basin then goes as $(\phi_J-\phi)$. The strictly-$1d$ hard rod system has an extremely simple inherent structure landscape consisting of a single basin that jams in a state with all the rods in contact end-to-end and $\phi_J=1$. The inherent structure landscape is more complicated in higher dimensions. In two dimensions, a disc is locally locally jammed if it has at least three rigid contacts, not all in the same hemisphere. However, a configuration made up of locally jammed discs is not necessarily a mechanically stable inherent structure, or collectively jammed state, because a collective motion of the particles can lead to unjamming~\cite{Torquato_Multiplicity_2001}, highlighting the global nature of jamming phenomena.
 
\begin{figure}[t!]
\includegraphics[width=1.0\linewidth]{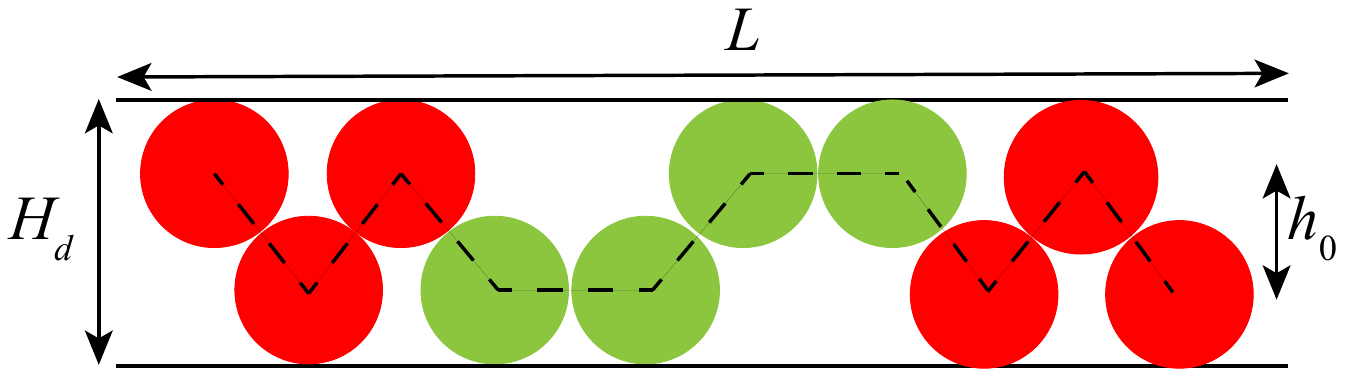}
\caption{A configuration of hard discs in a jammed packing containing discs in the most dense (red) and defect (green) local structures.}
\label{fig:sketch}
\end{figure}

Confining the particles within narrow quasi-$1d$ channels reduces the number of possible disc-disc contacts and prevents collective rearrangements of the particles. As a result, the complete inherent structure landscape of the current model, and how it is sampled as function of $\phi$, can be constructed by considering local packing arrangements~\cite{Yamchi:PRL2012,Ashwin:PRL2013,Yamchi:PRE2015}. The most dense structure consists of a zig-zag packing of discs contacting their two neighbours across the channel as well as the wall and the lower density jammed states are formed by introducing defects where discs form a contact with another disc along the channel (see Fig.~\ref{fig:sketch}). The jamming density of a packing is then given by~\cite{Bowles_Landscapes_2006},
\begin{equation}
	\phi_{J} = \frac{\pi}{4 H_{d}\left(\theta+\left(1-\theta\right)\sqrt{H_{d}\left(2\sigma-H_{d}\right)}\right)}\mbox{,}\\
	\label{eq:eqn3}
\end{equation}
where $\theta$ is the fraction of defects so that $\phi_{J}\rightarrow\phi_{J_{max}}$ as $\theta\rightarrow 0$. Noting that two defects cannot appear next to each other in a jammed configuration because the particles would not satisfy the local jamming condition, it has also been shown that the lowest density jammed structure consists of alternating dense and defect states, with $\theta=0.5$. The number of inherent structure basins is also a function of $\theta$, with the maximum number of jammed structures occurring when $\theta=1/2-\sqrt{5}/10$ and decreasing to unity as $\theta$ approaches its upper and lower limits.

At equilibrium, the fluid samples the set of basins on the inherent structure landscape that maximizes the total entropy so that at a given $\phi$ the system samples basins with the equilibrium fraction of defects $\theta=\theta^*$. The insert to Figure~\ref{fig:EOS} shows $\theta^*$ as function of $\phi$, obtained by maximizing the total entropy of the fluid~\cite{Ashwin:PRL2013,Yamchi:PRE2015}. At low $\phi$, the system samples the region of the inherent structure landscape where there is a large number of basins with a high defect fraction and low $\phi_J$. With increasing $\phi$, the system samples a smaller number of basins, and smaller defect fraction, that have a greater vibrational entropy due to their larger $\phi_J$, effectively trading a reduction of configurational entropy for increased vibrational entropy.

\begin{figure}[t!]
\centering
\includegraphics[width=1.0\linewidth]{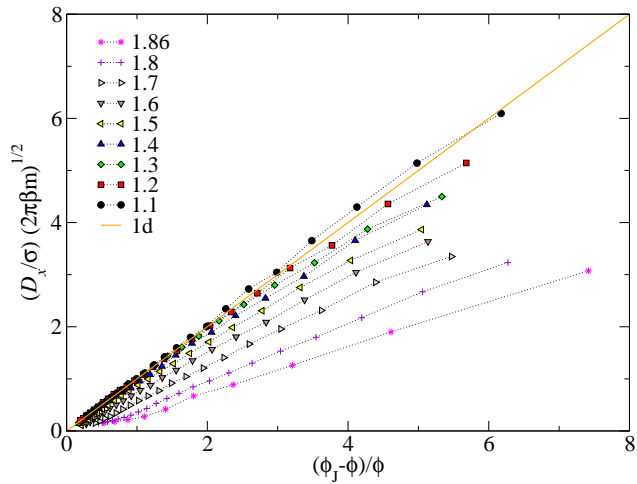}
\caption{Test of Eq.~\ref{eq:dx2d} plotting $(D_x/\sigma_N)(2 \pi \beta m)^{1/2}$ as a function of $(\phi_{J}-\phi)/\phi$.}
\label{fig:figure5}
\end{figure}

In the context of the current problem, this means that $\phi_J$ is a function of density, which in turn influences the amount of space the particles have move. 
 Equation.~\ref{eq:eqn2} can then be rewritten to account for density dependent jamming density as,
\begin{equation}
\frac{D_x}{\sigma} = \frac{(\phi_J-\phi)}{\phi (2 \pi \beta m)^{1/2}},
\label{eq:dx2d}
\end{equation}
where $(\phi_J-\phi)/\phi$ is the effective mean clearance between adjacent particles, and $\phi_J$ at each density is given by Eq.~\ref{eq:eqn3} using $\theta^*$. Figure~\ref{fig:figure5} shows the predictions of the Eq.~\ref{eq:dx2d} for different channel widths, where we have used the equilibrium value of $\theta$ to obtain $\phi_J$. While the data is somewhat linearized the slopes of the lines decrease with increasing $H_d$, which may be a result of the increased motion of the particles perpendicular to the longitudinal axis of the channel that is not accounted for in our measurements of the MSD.

The exact EOS for quasi-$1d$ systems was solved by Kofke and Post~\cite{Kofke:1993p3} using a transition matrix approach. Their approach uses the idea that when the {\it y} positions of the particles are fixed the system can be characterized as  a $1d$ mixture of hard rods on a line with different contact lengths. This allow the integration in $x$ to be performed independently of the {\it y} integration when solving the partition function. The results of the integrals can be solved as an eigenvalues problem where the largest eigenvalue is used, which essentially identifies an average effective diameter of the projected hard rods. 

\begin{figure}
\centering
 \includegraphics[width=1.0\linewidth]{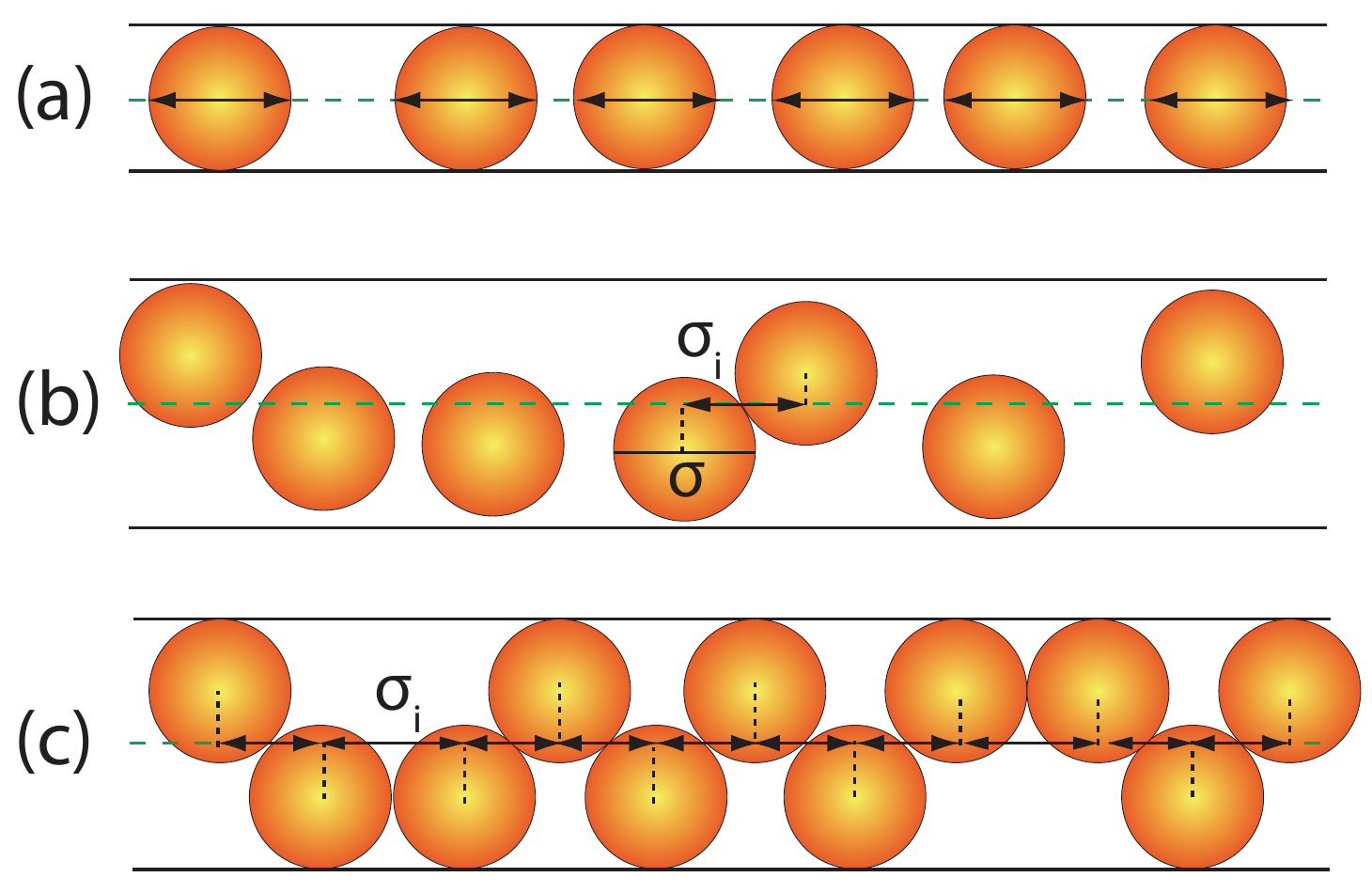}
   \caption{Effective diameter, a) purely $1d$ system $\sigma = \sigma_{N}$, b) $H_{d}=1.8$ at low density and c) $H_{d}=1.8$ at high density }
   \label{fig:figure6}
\end{figure}

Motivated by this, we define an effective one-dimensional interaction diameter, $\sigma_i$, between two neighbouring particles that simply depends on their relative separation in the $y$-direction, $\Delta y=|y_{i+1}-y_i|$,
\begin{equation}
\sigma_i=\sqrt{\sigma^2-\Delta y^2}\mbox{,}\\
\label{eq:edia}
\end{equation}
and reduces to $\sigma$ when $\Delta y=|y_{i+1}-y_i|=0$ (see Fig.~\ref{fig:figure6}).

The average effective $1d$ interaction diameter is then given by,
\begin{equation}
 \sigma_{\textup{N}} =\frac{1}{N}\sum_{i} \sigma_{\textup{i}},
 \label{eq:eqn4}
\end {equation}
and we can define an effective occupied volume fraction,
\begin{equation}
\phi_{\textup{eff}}=\frac{N\sigma_{\textup{N}}}{L}\mbox{,}\\
\label{eq:eqn5}
\end{equation}
that can be extracted from the equation of state of the quasi-one dimensional system and used to obtain the diffusion coefficient from Eq~\ref{eq:eqn2}. To achieve this, we assume the exact $1d$ equation of state holds for $\phi_{\textup{eff}}$:
\begin{equation}
Z = \frac{1}{1-\phi_{\textup{eff}}}  ,
\label{eq:eqn6}
\end{equation}
where $Z=PV/NkT$ is the compressibility factor for the quasi-one-dimensional system. Rearranging, we obtain $\phi_{\textup{eff}}$ as,
\begin{equation}
\phi_{\textup{eff}} = \frac{Z-1}{Z} \mbox{.}\\
\label{eq:eqn7}
\end{equation}
Sububstituting eq.~\ref{eq:eqn5} into ~\ref{eq:eqn2}, gives the diffusion coefficient in terms of the effective volume fraction,
\begin{equation}
\frac{D_{x}}{\sigma_\textup{N}} = \frac{(1-\phi_{\textup{eff}})}{\phi_{\textup{eff}} (2 \pi \beta m)^{1/2}}\mbox{.}\\
\label{eq:eqn8}
\end{equation}

\begin{figure}
\centering
 \includegraphics[width=1.0\linewidth]{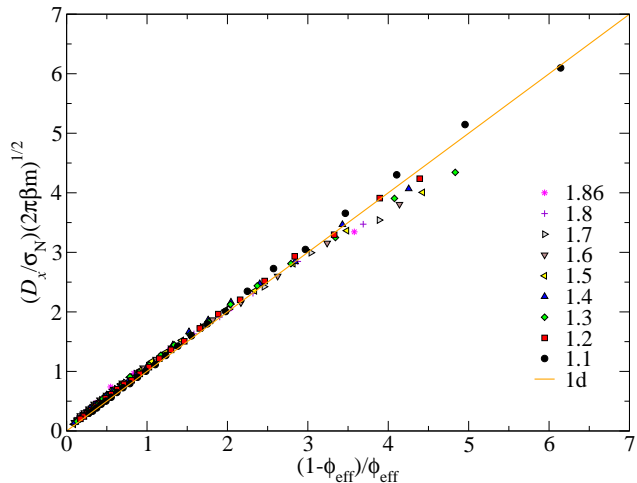}
  \caption{Scaled self-diffusion coefficient data represented by using Eq.~\ref{eq:eqn8} by using effective volume fraction for various channel width.}
   \label{fig:figure7}
\end{figure}

Figure~\ref{fig:figure7} shows Eq.~\ref{eq:eqn8} using the simulated data for the diffusion coefficient at different channel widths compared to exact analytical results for $1d$ system. The plot shows, the collapse of the data is good,  
which suggests $\phi_{\textup{eff}}$ provides a good thermodynamic connection through the equation of state to $1d$ diffusion, as suggested by Percus~\cite{percus:2010jsp}. The deviation at low densities may be related to a broad distribution of effective diameters at low densities, where the particles move freely in the transverse direction and can adopt many different effective diameters. At high densities, the distribution of effective diameters is restricted more closely to the average, except in the cases where the particles are appear in defect states and $\sigma_i\approx \sigma$. In addition, the measurements of $D_x$ at the very lowest densities can be difficult because of the low frequency of particle-particle collisions and the slow convergence of $\gamma\rightarrow1$.

\begin{figure}
\centering
 \includegraphics[width=1.0\linewidth]{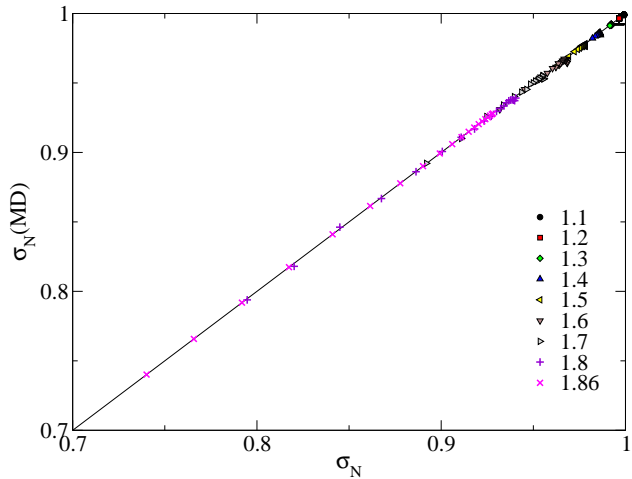}
   \caption{Effective diameter calculated from Eqs. ~\ref{eq:eqn5} and ~\ref{eq:eqn6} vs. those calculated from MD simulations using Eq.~\ref{eq:edia} for different $H_d$.}
   \label{fig:Effectivediam}
\end{figure}

While $\phi_{\textup{eff}}$ is obtained from the equation of state, it can be measured geometrically. Figure~\ref{fig:Effectivediam} compares the values of $\sigma_{N}$ obtained using eq.~\ref{eq:eqn6} with the average values of the effective diameter measured during our MD simulations at different channel widths. This shows the usefulness of this quantity especially, in the cases that the exact EOS is not available.

\begin{figure}
\includegraphics[width=1.0\linewidth]{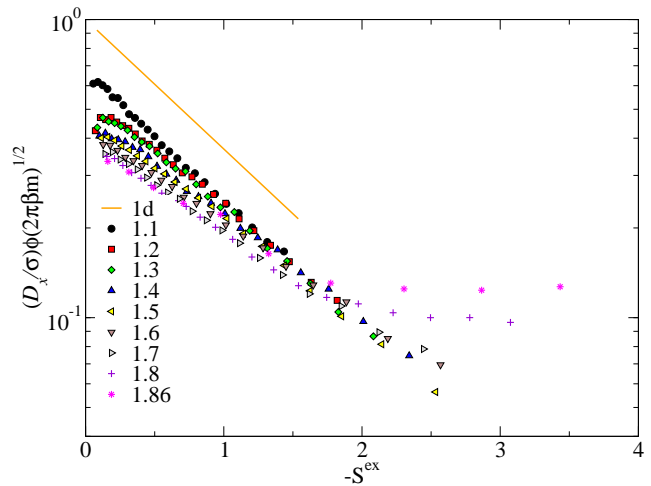}
\caption{Scaled self-diffusion coefficient data versus negative excess entropy calculated using Eq.~\ref{eq:eqn9} for different $H_d$.}
\label{fig:figure8}
\end{figure}

Finally, we simply demonstrate that $\phi_{\textup{eff}}$ also provides an appropriate link between the excess entropy and diffusion in these confined fluid systems. 
Here, the excess entropy of a strictly one-dimensional hard rod system, relative to ideal gas, can be obtained,
\begin{equation}
s^{\textup{ex}}/Nk  = -\int_{0}^{\phi} \frac{Z-1}{\phi^{'}}\textup{d} \phi^{'} ,
\label{eq:eqn9}
\end{equation}
which yields,
\begin{equation}
s^{\textup{ex}}/Nk  = \ln(1-\phi)\mbox{.}\\
\label{eq:eqn10}
\end{equation}
Again, replacing $\phi$ with $\phi_{\textup{eff}}$ in Eq. \ref{eq:eqn10}, and using Eq.~\ref{eq:eqn8} yields,
\begin{equation}
D_x= \left(\sigma_{N}/\phi_{\textup{eff}}\right) (2 \pi \beta m)^{-1/2}\textup{exp}\left (s^{\textup{ex}}\right )\mbox{,}
\label{eq:eqn11}
\end{equation}
which connects the excess entropy to a reduced diffusion coefficient, $D_x\phi_{\textup{eff}}/\sigma_{N}$, in a similar form to that obtained by Rosenfeld~\cite{Rosenfeld:1977p14198} and Dzugutov~\cite{Dzugutov:1996p14200}. Figure~\ref{fig:figure8} shows the relationship between the diffusion coefficient and excess entropy in the quasi-one-dimensional system before rescaling. Then use of  $\phi_{\textup{eff}}$ necessarily reproduces the same collapse observed in Fig.~\ref{fig:figure7}.

\section{Conclusion}
\label{sec:con}
Simulation studies of both bulk and confined fluids have shown the effectiveness of scaling relationships that map the transport properties of the system to a reference state through a thermodynamic property such as the excess entropy. In the case of highly confined, quasi-one-dimensional fluids, the strictly one-dimensional fluid is an ideal reference state because results are often known exactly. This work shows that the proposed scaling can be achieved for a system of quasi-one-dimensional hard discs through an effective occupied volume fraction that is obtained directly from the equation of state or can be measured geometrically. However, the channel widths considered here are still narrow and it is likely the scaling will begin to breakdown for wider channel. It also remains to be seen how well the scaling works for longer range interactions.

\begin{acknowledgments} 
We thank NSERC for financial support. We also thank WestGrid and Compute Canada for providing computational resources.
\end{acknowledgments}

%
\bibliography{ref.bib}
%



\end{document}